\documentclass[aps,prd,showpacs,11 pt]{revtex4-2}

\usepackage{slashed}
\usepackage{mathtools}
\usepackage{hyperref}
\usepackage{amsmath}
\usepackage{amssymb}
\usepackage{latexsym}
\usepackage{amsfonts}
\usepackage{epsfig}
\usepackage{psfrag}
\usepackage{graphicx}
\usepackage{setspace}
\usepackage{braket}
\usepackage{color}
\def\hhref#1{\href{http://arxiv.org/abs/#1}{arXiv:#1}} 
\usepackage{siunitx}

\begin{document}

\title{QED vacuum nonlinearity in Laguerre-Gauss beams}

\author{Cesim K. Dumlu, Yoshihide Nakamiya and Kazuo A. Tanaka}

\affiliation{Extreme Light Infrastructure-Nuclear Physics (ELI-NP), 077125, M\u{a}gurele, Romania}

\email{cesim.dumlu@eli-np.ro}

\begin{abstract}
We investigate in detail the stimulated photon emission in pulse-shaped Laguerre-Gauss beams of arbitrary mode decomposition within the framework of Euler-Heisenberg effective theory. We elaborate on the analytic structure of the photon emission amplitude and identify its dominant and sub-dominant parts by analyzing each signature's phase scaling.  We identify three distinct nonlinear signatures: the orbital angular momentum (OAM) flip ($\ell:1 \rightarrow -1$) , higher harmonic generation ($\ell:1 \rightarrow 2$)  and harmonic reduction ($\ell:1 \rightarrow 0$). We  estimate  signal photon yield  for each signature under various pulse collision scenarios we explain in detail.  We also outline a prospective experimental setup that may probe these signatures.     

\end{abstract}


\maketitle


\section{Introduction}
     
Euler-Heisenberg Lagrangian encodes the self-coupling of the electromagnetic fields, giving rise to various nonlinear effects of quantum origin \cite{he,schw,dun1}. This coupling originates from the  photon-photon interaction \cite{kn},  the  evidence for which was reported by the ATLAS collaboration in heavy ion collisions\cite{aaboud}. The direct observation of the photon scattering based on a beam collision scenario however remains to be a challenging task, ideally requiring a collision between ultra-brilliant gamma beams to overcome the detection threshold; yet controlled gamma-gamma collisions appear to be beyond the reach of currently available technology (see \cite{sz} and references therein). The effect of these nonlinear couplings can on the other hand  be seen in various phenomena, (see \cite{rev1} and \cite{rev2} for a review) including vacuum birefringence \cite{vb1,vb2,vb3,vb4,vb5,vb6,vb7, vb8}, photon splitting\cite{ps1,ps2,ps3,ps4,ps5} and  stimulated photon emission \cite{pe1,pe2,pe3,pe4,pe5,pe6,pe7}. The experimental verification of these phenomena could be in near future, given the significant advancements in the laser technology \cite{ls1,ls2,ls3,ls4,ls5,ls6,ls7}. 

Here we are interested in stimulated photon emission process, during which several quanta from the external field can be absorbed and emitted as signal photons \cite{pe01,pe02,pe03}. These signal photons can be generated at higher harmonics, which is the telltale sign for the nonlinear coupling of the external macroscopic fields \cite{hh1,hh2,hh3}.  Inside the focal area where two counter-propagating Gaussian pulses overlap, the signal photon yield roughly scales as: $\sim \alpha (E_0/E_s)^6 (w_0/\lambda_{C})^4 $ \cite{ks}. Here, $\alpha$ is the fine structure constant, $E_s$ is Schwinger critical field and $\lambda_C$ is the reduced Compton wavelength.  This simple scaling relation shows that at high field strength, $E_0$ and with large focal spot size $w_0$, significant amount of signal quanta can be induced by the external fields. An early estimate on the signal photon production rate was reported in \cite{dew}, based on the far field solutions of nonlinear Maxwell equations \cite{nm1,nm2,nm3,nm4,nm5}. Later studies, using the far field approach,  have shown that the signal photon yield can reach up to $10^3-10^4$ in a collision between two or three Petawatt-class laser beams \cite{pe1, pe2}. In the alternative treatments the photon yield was found by directly computing the Fourier transform of the nonlinear couplings \cite{ks, khs}.

The search for these signal photons is anticipated to be a formidable task to say the least. Among the many challenges involved, one problem particularly relevant for the recent studies is the enhancement of signal to background ratio; the signal photons are largely emitted along the forward cones of the driving pulses and this makes it challenging to sort them out from the background photons. The characteristics  of signal photons such as their polarization, frequency and angular spectrum play an important role in differentiating them from the background; in this respect, non-planar, multi-pulse collision setups have been the focus of the several works \cite{pe1, khs, karb1}. In the collision scenarios, where  the driving pulses  are prepared in second and higher order harmonics,  signal  photons at several distinct higher harmonics can be induced with enhanced far-field amplitudes \cite{karb2}.  A novel approach based on the far-field characteristics Laguerre-Gauss beams was  given in \cite{km1}. Recently in \cite{orb}, the authors have considered a planar, three-beam collision scenario with a twist: in addition to the energy signature, it was shown that orbital angular momentum(OAM) of light can used as an additional filter that may distinguish the signal photons from the background. Working in the framework of nonlinear Maxwell theory, the authors have used a Bessel beam (with $\ell=1$) as one of the external fields, and have estimated the number of signal photons emitted with the flipped OAM: $\ell=-1$.

Here, in view of the idea developed in \cite{orb}, we explore the photon emission signatures by using Laguerre-Gauss (LG$_{\ell \, p}$) beams  as the driving pulses. LG$_{\ell\,p}$ beams are the solutions of Maxwell equations in the paraxial limit and  they are the natural extension of the Gaussian beam to higher order angular and radial modes. Compared to Bessel modes, LG$_{\ell\,p}$ modes have robust propagation properties \cite{lg1} and have a wide variety of potential use cases, some of which include optical traps in Bose-Einstein condensates \cite{lguc1}, electron acceleration \cite{lguc2}, gravitational wave interferometry\cite{lguc3} and quantum entanglement\cite{lguc4,lguc5}.  In the following, we work with the method of \cite{ks} and obtain  formulas for the photon emission amplitudes for LG$_{\ell\,p}$ beams of arbitrary mode decomposition. In particular, we analyze in detail three distinct nonlinear signatures: the OAM flip ($\ell:1\rightarrow -1 $),  higher harmonic generation ($\ell: 1\rightarrow 2$), which we shall refer as the doubling signature, and finally the harmonic reduction ($\ell:1 \rightarrow 0$).  We also elaborate on a prospective experimental scenario that may probe such signatures. The organization of this paper is as follows. In Section II, we briefly revisit the theoretical aspects and analyze in detail the analytic structure of the photon emission amplitude. In Section III, we discuss the pulse collision scenarios along with the aforementioned signatures and present our results for the signal photon yield for each case. Final section contains our conclusions.

\section{The Theory}

Here, we briefly outline the theoretical aspects of the approach, the details of which were elucidated in the works \cite{ks, pe6} (see also \cite{pefurry} for Furry picture analysis). We are interested in  the transition amplitude:
\begin{eqnarray}
S_{p}(\vec{k})= \frac{i}{\hbar}\langle \, \gamma_{p}(\vec{k}) | \hat{\Gamma}^{1}_{\text{int}}[A_{\text{ext}}]\, | 0 \rangle
\end{eqnarray}
where
\begin{eqnarray}
\hat{\Gamma}^{1}_{\text{int}}[A_{\text{ext}}]=\int d^4 x\, \delta \Gamma^{1}_{\text{EH}}[\hat{A}]\big|_{A=A_{\text{ext}}}
\end{eqnarray}
is the operator encoding the transition to the final state:  $\gamma_{p}(\vec{k})\rangle\equiv a^{\dagger}_{\vec{k},\,p}|0\rangle$, which is labeled by the polarization $p$ and momentum $\vec{k}$  of the outgoing (quantized) signal photon.  The operator $\delta\Gamma^{1}_{\text{EH}}[\hat{A}]\big|_{A=A_{\text{ext}}}$ acts on the signal photon Fock state  after the total field is split into the external (classical) and signal parts: $A\rightarrow A_{\text{ext}}+\hat{a}$ (we refer reader to \cite{ks2} and \cite{gk} for further details on this splitting procedure), and  the  variation is performed at $\hat{A}=A_{\text{ext}}$ up to the linear order in the quantized field. This ultimately yields: $\delta\Gamma^{1}_{\text{EH}}[A_{\text{ext}}]\equiv \partial \mathcal{L}^1_{\text{EH}} /\partial F^{\mu\nu}_{\text{ext}} \hat{f}^{\mu\nu} $, where $\mathcal{L}^1_{\text{EH}}$ is the one-loop Euler-Heisenberg Lagrangian. Here,  $\hat{f}^{\mu\nu} =\partial^{\mu}\hat{a}^{\nu}-\partial^{\nu}\hat{a}^{\mu}$ and
\begin{equation}
\hat{a}^{\mu}=\sum_{p=1,2} \int \frac{d^3 k}{(2\pi)^3} \sqrt{\frac{\hbar c^2}{2 \omega \epsilon_0}} \left(\epsilon^{\mu}_{p}(\vec{k}) e^{- i k x} \hat{a}_{\vec{k},\, p} +\epsilon^{*\,\mu}_{p}(\vec{k}) e^{i k x} \hat{a}^{\dagger}_{\vec{k},\, p} \right)
\end{equation}
is the signal field,  with the momentum $ k^{\mu}:(\omega/c ,\, \vec{k}), \, k^{2}=0$ and it  is spanned  by the space-like polarization vectors ($\epsilon^{\mu}_{p}(\vec{k}), \, \epsilon^{*\,\mu}_{p}(\vec{k})$),  which are given in the Lorenz gauge. To proceed, we insert (3) into (1) and use the leading order $ \mathcal{L}^1_{\text{EH}}$. Doing so the transition amplitude reads ($g_{\mu\nu}:=-+++$):  
\begin{eqnarray}
S_{p}(\vec{k})=\frac{1}{4\pi}\frac{e^4}{45 \pi m^4 c^7} \sqrt{\frac{\hbar \omega}{2 \epsilon_0}}\int d^4 x \left[4\mathcal{F} \left(\vec{\epsilon}^{\,\,*}_{p}.\vec{E} - (\hat{k}\times \vec{\epsilon}^{\,\,*}_{p}).\vec{H}\right)+ 7\mathcal{G}\left(\vec{\epsilon}^{\,\,*}_{p}.\vec{H} + (\hat{k}\times \vec{\epsilon}^{\,\,*}_{p}).\vec{E} \right)\right] e^{ i k x}
\end{eqnarray}    
where we have used the  definition: $\vec{B} c=\vec{H}$  and the field invariants are defined as: $\mathcal{F}=\frac{1}{4} F_{\mu\nu}F^{\mu\nu}=\frac{1}{2}(\vec{H}^2  -\vec{E}^2)$ and $\mathcal{G}=\frac{1}{4}F_{\mu\nu}\prescript{*}{}{F^{\mu\nu}}=-\vec{E}.\vec{H}$ ,where $\prescript{*}{}{F^{\mu\nu}}=\frac{1}{2}\epsilon^{\mu\nu\alpha\beta}F_{\alpha\beta}$. The amplitude $S_{p}(\vec{k})$ depends on the characteristics of the background field, which is assumed to vary slowly over the length scale $\sim\lambda_{C}$. This means the external fields  in the optical regime with femtosecond order pulse durations can be treated as locally constant, justifying the use of  $ \Gamma^{1}_{\text{EH}}$ in the variational approach outlined above. The  total unpolarized particle number is given by
\begin{eqnarray}
N= \frac{1}{(2\pi)^3}\int d^3k \sum_{p=1,2} |S_{p}(\vec{k})|^2
\end{eqnarray}
where summation runs over the possible polarization states of the outgoing photon.

In the remainder of this paper, we will focus on two-beam head on collision scenario, in which the first and second beams respectively travel along $\hat{z}$ and $-\hat{z}$. In such a setting,  the phase of the external fields becomes separable and the transition amplitude has symmetry around the optical axis. This means all the components of the Fourier integrals in (4), apart from $z$-dependent one, can be performed analytically. To proceed further, we assume that external fields are paraxial, characterized by a single global polarization vector and satisfy ($i=1,2$): $\vec{E}_{i}. \vec{H}_{i} =0 $ and $|\vec{E}_{i}|=|\vec{H}_{i}|$. Taking these into account  and choosing $\vec{E}_1=E \,\hat{x}$ and $\vec{E}_2=-\bar{E}\, \hat{x} $  completely fixes the vector parametrization of the external fields. For the signal photon's  momentum  we use the parametrization: $\hat{\vec{k}}= (\cos{\phi_{k}}\sin{\theta_{k}},\, \sin{\phi_{k}}\sin{\theta_{k}},\,   \cos{\theta_{k}} ) $ and denote its polarization vector as
\begin{eqnarray}
\vec{\epsilon}_1  = 
  \begin{bmatrix}
    \cos{\phi_{k}}\cos{\theta_{k}}\cos{\beta}-\sin{\phi_{k}}\sin{\beta}\\
   \sin{\phi_{k}}\cos{\theta_{k}}\cos{\beta}+\cos{\phi_{k}}\sin{\beta} \\
   -\sin{\theta_{k}}\cos{\beta} 
   \end{bmatrix} 
\end{eqnarray}
where the remaining polarization vector is simply given by: $\vec{\epsilon}_2 (\beta)= \vec{\epsilon}_1 (\beta+\pi/2)$. The value of $\beta$ here is  arbitrary and can be fixed along the desired direction, with respect to which the polarization axes of the signal photon are defined. In the following, we will concern ourselves with the total yield of the signal photons so the choice of $\beta$ will not be relevant. We may write the final form of the signal photon number as
\begin{equation}
N= \frac{1}{2\pi^3}\int |\vec{k}|^3 \, d|\vec{k}| \sin{\theta_k} \, d\theta_k \, d\phi_k \, n(\vec{k})
\end{equation}
with
\begin{eqnarray}
n(\vec{k})&=& 256\mathcal{P} \bigg(\sin^4 \frac{\theta_k}{2}  |\mathcal{S}(\vec{k})|^2 +\cos^4 \frac{\theta_k}{2}  |\mathcal{\bar{S}}(\vec{k})|^2  \bigg) \nonumber\\
&+& 256\mathcal{P} \sin^2 \frac{\theta_k}{2} \cos^2 \frac{\theta_k}{2} \cos{2\phi_k}\bigg(\mathcal{S}(\vec{k})^* \mathcal{\bar{S}}(\vec{k})+\mathcal{S}(\vec{k}) \mathcal{\bar{S}}^*(\vec{k})\bigg)
\label{pdens}
\end{eqnarray}
where the prefactor is:
\begin{eqnarray}
\mathcal{P}= \frac{\alpha}{2\pi^3}\frac{c^2}{90^2}\frac{1}{\lambda_C^4 E_s^6},\quad \lambda_C=\frac{\hbar}{mc},\,\, E_s=\frac{m^2 c^3}{e \hbar}
\end{eqnarray}
and the Fourier amplitudes are:
\begin{equation}
\mathcal{S}(\vec{k})= \int d^4 x \, e^{i k x} E^2 \bar{E}, \quad  \mathcal{\bar{S}}(\vec{k})= \int d^4 x \, e^{i k x} E \bar{E}^2
\label{famp}
\end{equation}
As we will shortly explain, above respectively represent backward and forward emission amplitudes with respect to the optical axis.  $E$ and $\bar{E}$ are the electric fields propagating respectively along  $\hat{z}$  and $-\hat{z}$ and  they are modeled as real part of the LG$_{\ell\,p}$ beam:
\begin{eqnarray}
E &=& \sum_{\ell\, p}E^{\ell\, p}_0 \frac{w_0}{w(z)}\left(\dfrac{\sqrt{2} r}{w(z)}\right)^{|\ell|} \cos\left(\Omega (z/c-t) -\ell \phi + \phi_{\ell\, p} - (2p +|\ell|+1)\text{tan}^{-1}\frac{z}{z_R} + \frac{z}{z_R}\frac{r^2}{w(z)^2}\right)   \nonumber\\& \times & e^{-\frac{r^2}{w(z)^2}} L^{|\ell|}_{p}\left(\frac{2 r^2}{w(z)^2}\right)e^{-4(z/c - t)^2/\tau^2}
\label{ef}
\end{eqnarray}
with waist size  $w(z)=w_0\sqrt{1+ z^2/z^2_R}$,  central frequency $\Omega$, Rayleigh length $z_R$ and pulse duration $\tau$. The mode-dependent quantities are the phase off-set $\phi_{\ell\,p}$ and the field strength $E^{\ell\,p}_0$, whose definition will be given in the next section. Here $L^{|\ell|}_{p}$ represents the associated Laguerre polynomial. Note that the term $\sim r^{|\ell|}$ vanishes at origin, encoding the topology of the beam with a non-integrable phase factor, which is quantized by the integer values of $\ell$ (also referred as the topological charge).  For further details on the characteristics of LG$_{\ell,p}$ beam we refer the reader to \cite{lg2}. Counter-propagating pulse $\bar{E}$ is defined via the replacement $z\rightarrow -z$ and has independent parameters which shall be denoted by the barred notation: $E^{\bar{\ell},\,\bar{p}}_{\bar{0}}$,  $\bar{z}_R$, $\bar{\Omega}$ and so on.  In (\ref{ef}) we have  introduced a Gaussian damping factor  in the longitudinal domain  to simulate the fact that physical pulses carry finite energy.  This specific form in fact follows from the broadband character of the pulse, for which we have assumed  here to be Gaussian, $\sim e^{-(\kappa-\kappa_o)/\Delta^2}$ with a central wavenumber $\kappa_0 =  \Omega/c $ and bandwidth $\Delta=\kappa_{\text{max}}-\kappa_{\text{min}}$. The pulse duration is related to the bandwidth as $\tau \sim  \Delta  / c \kappa_{\text{max}}\kappa_{\text{min}} $. Taking these into consideration, the longitudinal profile of the pulse $\sim e^{i \kappa (c t- z)}$ when convoluted with such distribution indeed leads to the Gaussian damping factor as given above. We should note that the above form assumes temporal and spatial components of the pulse are separated; in high power laser systems however spatio-temporal couplings can occur during the propagation \cite{stc1,stc2}. Nevertheless, the fuller treatment of the pulse shape is beyond the scope of the present work and we will content ourselves here with the form given in (\ref{ef}).  It is also worth mentioning that such damping factor will introduce  additional terms to the paraxial Helmholtz equation. For a Gaussian pulse the paraxiality condition at the pulse center is given by the inequality $4/w_0^2 \kappa^2 \ll 1$. When the damping factor included this condition now becomes  $4/w_0^2 \kappa^2 + w_0^2/c^2\tau^2 \ll 1$.  For higher order modes the paraxiality condition reads \cite{vrl} : $(2p + |\ell|+1)/\kappa^2 w_0^2 \ll 1$. These inequalities are satisfied for the parameter regime of our interest, for which we will consider pulse durations of $\tau \gtrsim 10$ femtoseconds, focal radius of $w_0 \lesssim 5$ microns, and  the low order modes: $\ell=1,0,\, p=0$.

In the following we elaborate on the analytic structure of the emission amplitude.  The Fourier amplitudes in (\ref{famp}) contain cubic powers of cosines each of which contains negative and positive frequency parts, leading to a total of 8 sign permutations. We perform the integration of (\ref{famp}) over each permutation, after separating  $ E^2 \bar{E}$ and $E \bar{E}^2$ into their respective phase and amplitude components. The advantage of doing so is that it makes the energy and OAM signature of the signal obvious. In addition, as we will shortly discuss, it turns out that this way one can easily identify the parts of the amplitude which are exponentially suppressed. Doing so, the Gaussian integrals over time and radial directions can easily be performed. The angular integral can also be integrated by making use of a well-known identity, whereas the remaining $z$ dependent integral is to be computed numerically.  The final form of the Fourier amplitudes are given as
\begin{eqnarray}
\mathcal{S}(\vec{k})&=&\frac{1}{8} \sum_{\substack{\ell_1,\, \ell_2,\,\bar{\ell}_3,\\ p_1,\,p_2,\,\bar{p}_3 }}\sum_{\substack{s_1,\,s_2,\,s_3}} S^{s_1 s_2 s_3}_{\{\ell,\, p \}}, \quad \quad 
\mathcal{\bar{S}}(\vec{k})=\frac{1}{8} \sum_{\substack{\ell_1,\, \bar{\ell}_2,\,\bar{\ell}_3,\\ p_1,\,\bar{p}_2,\,\bar{p}_3 }} \sum_{\substack{s_1,\,s_2,\,s_3}} \bar{S}^{s_1 s_2 s_3}_{\{\ell,\, p \}}
\label{ampsum}
\end{eqnarray}
where each $s_{i}=\pm$ labels positive and negative frequency parts and  $\{\ell,\, p \}$ collectively denotes the quantum numbers. The summands are
\begin{eqnarray}
S^{s_1 s_2 s_3}_{\{\ell,\, p \}}  &=& E_0^{\ell_1,\, p_1}E_{0}^{\ell_2,\, p_2}E_{\bar{0}}^{\bar{\ell}_3,\,\bar{p}_3}  \bigg(\frac{2\pi z^2_{R} c}{\Omega}\bigg) \bigg(\frac{\pi^\frac{1}{2}\tau}{2\sqrt{\tau^2_{12}+2}}e^{-\frac{\tau^2}{16(\tau_{12}^2+2)} (|\vec{k}|c-(s_1+s_2+s_3\Omega_{21})\Omega)^2} \bigg)   \nonumber\\
&\times & e^{i (s_1 \ell_1 + s_2\ell_2 +s_3 \bar{\ell}_3) \phi_k} \mathcal{F}^{s_1 s_2 s_3}_{\{\ell,\, p \}}\nonumber\\
\bar{S}^{s_1 s_2 s_3}_{\{\ell,\, p \}}  &=& E_0^{\ell_1,\, p_1}E_{\bar{0}}^{\bar{\ell}_2,\, \bar{p}_2}E_{\bar{0}}^{\bar{\ell}_3,\,\bar{p}_3}  \bigg(\frac{2\pi z^2_{R} c}{\Omega}\bigg) \bigg(\frac{\pi^\frac{1}{2}\tau}{2\sqrt{2\tau^2_{12}+1}}e^{-\frac{\tau^2}{16(2\tau_{12}^2+1)} (|\vec{k}|c-(s_1+(s_2+s_3)\Omega_{21})\Omega)^2} \bigg)   \nonumber\\
&\times & e^{i (s_1 \ell_1 + s_2\bar{\ell}_2 +s_3 \bar{\ell}_3) \phi_k} \mathcal{\bar{F}}^{s_1 s_2 s_3}_{\{\ell,\, p \}}
\label{amp}
\end{eqnarray}     
The terms $\mathcal{F}^{s_1 s_2 s_3}_{\{\ell,\, p \}}$ and $\mathcal{\bar{F}}^{s_1 s_2 s_3}_{\{\ell,\, p \}}$ are dimensionless and encode  the remaining integrals along $z$ direction. We give the details of the integration and the full result in the Appendix.  The ratios $\tau_{12}$ and $\Omega_{21}$ are  defined as: $\tau_{12}=\tau/\bar{\tau}$ and $\Omega_{21}=\bar{\Omega}/\Omega$. There is also a third ratio: $z^{21}_{R}=\bar{z}_R/z_R$ embedded  in the integrals along $z$.   Looking closely at (\ref{amp}) we can readily identify $2\pi z^2_{R} c/\Omega$ as the volume factor. The remaining factors inside the parentheses are the result of the time integrals and determine the energy signature, which in the formal limit: $\tau\rightarrow \infty$, reduce to $2\pi\delta$ with the corresponding arguments. The important point here is that the dominant contribution to signal photon yield comes respectively from the regions where $|\vec{k}| \sim (s_1+s_2+s_3\Omega_{21})\Omega/c$ and $|\vec{k}|\sim(s_1+(s_2+s_3)\Omega_{21})\Omega/c$.  The remaining exponential terms represent the total OAM signature of the emitted photon and it is related to  the energy signature, through $s_{i}$, and depends on the OAM of the photons partaking in the reaction.  Before proceeding, it is worth noting  that not all the signatures contribute to the photon yield.  For instance, when $\Omega_{21} \leq 2 $,  the negative energy signatures represented by the  amplitudes $S^{+--}_{\{\ell,\, p \}}, \,  S^{-+-}_{\{\ell,\, p \}}, \,  S^{--+}_{\{\ell,\, p \}},\, S^{---}_{\{\ell,\, p \}}$ and any cross terms between different energy signatures have vanishingly small support on $|\vec{k}| \in (0, \infty) $, therefore their contribution to the  photon yield effectively vanishes (and vanishes exactly in the formal limit,  $\tau \rightarrow \infty$).  The same arguments  apply to  the forward emission amplitude as well. 

In addition to this, the amplitudes have the following exchange symmetry
\begin{eqnarray}
S^{s_1 s_2 s_3}_{\{\ell_1 ,\ell_2, \bar{\ell}_3 , p_1,p_2, \bar{p}_3\}} &=& S^{s_2 s_1 s_3}_{\{\ell_2 ,\ell_1, \bar{\ell}_3,  p_2,p_1, \bar{p}_3\}} \nonumber\\
\bar{S}^{s_1 s_2 s_3}_{\{\ell_1 ,\bar{\ell}_2 ,\bar{\ell}_3 ,p_1,  \bar{p}_2,\bar{p}_3\}} &=& \bar{S}^{s_1 s_3 s_2}_{\{\ell_1,\bar{\ell}_3,\bar{\ell}_2 , p_1, \bar{p}_3 , \bar{p}_2\}}
\label{exsym}
\end{eqnarray}
We will make use of this property in the following section. Another observation that will be relevant for the rest of our discussion is the fact that certain signatures are suppressed. If we take a closer look  at the terms  $\mathcal{F}^{s_1 s_2 s_3}_{\{\ell,\, p \}}$  and  $\mathcal{F}^{s_1 s_2 s_3}_{\{\ell,\, p \}}$ ($\bar{z}=z/z_R$) :
\begin{eqnarray}
\mathcal{F}^{s_1 s_2 s_3}_{\{\ell,\, p \}}&=&\int^{\infty}_{-\infty}d \bar{z}\,  \text{exp}\bigg(-\frac{32 \bar{z}^2 z^2_{R}}{c^2 \tau^2(1+2/\tau_{12}^2)}\bigg) \text{exp}\big(i|\vec{k}|\bar{z}z_R\cos{\theta_k}-i\bar{z}z_{R}(s_1+s_2-s_3\Omega_{21})\Omega/c\big)\,...\nonumber\\
\mathcal{\bar{F}}^{s_1 s_2 s_3}_{\{\ell,\, p \}}&=&\int^{\infty}_{-\infty}d \bar{z}\,  \text{exp}\bigg(-\frac{32 \bar{z}^2 z^2_{R}}{c^2 \tau^2(2+1/\tau_{12}^2)}\bigg) \text{exp}\big(i|\vec{k}|\bar{z}z_R\cos{\theta_k}-i\bar{z}z_{R}(s_1-(s_2+s_3)\Omega_{21})\Omega/c\big)\, ...\nonumber\\
\label{zamp}
\end{eqnarray}
it becomes straightforward to see that the magnitude of the phase terms above is important because it determines how fast the above integrals damp. To illustrate, we fix $\Omega_{21}=1$ and recall that $|\vec{k}| \sim (s_1+s_2+s_3\Omega_{21})\Omega/c$. Then the overall phase in $\mathcal{F}^{+ + -}_{\{\ell,\, p \}} $  scales as $\sim i \bar{z} z_R\Omega/c  (\cos{\theta_k}-3)$ and in $\mathcal{F}^{+ + +}_{\{\ell,\, p \}} $ it becomes $\sim i\bar{z} z_R\Omega/c (3\cos{\theta_k}-1)$. Comparing these with  $\mathcal{F}^{+ - +}_{\{\ell,\, p \}} $   and $\mathcal{F}^{- + +}_{\{\ell,\, p \}} $, whose overall phase behave as $\sim i\bar{z} z_R\Omega/c  (\cos{\theta_k}+1)$, it becomes evident that the integrands in $\mathcal{F}^{+ + -}_{\{\ell,\, p \}} $ and  $\mathcal{F}^{+ + +}_{\{\ell,\, p \}} $ on the average  oscillate more rapidly and therefore  the values of these integrals become exponentially suppressed with respect to  $\mathcal{F}^{+ - +}_{\{\ell,\, p \}} $   and $\mathcal{F}^{- + +}_{\{\ell,\, p \}} $. 
From the scaling behavior noted above it becomes clear that the magnitude of the phase for  $\mathcal{F}^{+ - +}_{\{\ell,\, p \}} $   and $\mathcal{F}^{- + +}_{\{\ell,\, p \}} $ is minimized when  $\cos{\theta_k}\sim -1$. This establishes the fact that the amplitude $\mathcal{S}(\vec{k})$ has the dominant contribution in the backward scattering region.  The same observations also apply to the forward emission amplitude  where now integrands belonging to $\mathcal{\bar{F}}^{- + +}_{\{\ell,\, p \}} $ and $\mathcal{\bar{F}}^{+ + +}_{\{\ell,\, p \}} $ oscillate more rapidly  with respect to that of  $\mathcal{\bar{F}}^{+ - +}_{\{\ell,\, p \}} $ and $\mathcal{\bar{F}}^{+ + -}_{\{\ell,\, p \}} $, for which the magnitude of the overall phase is minimum when $\cos{\theta_k}\sim 1$. We should also note that the interference term given in the second line of (\ref{pdens}) is also suppressed. This is simply because it involves the cross terms $(S^{s_1 s_2 s_3}_{\{\ell,\, p \}})^*\bar{S}^{s_1 s_2 s_3}_{\{\ell,\, p \}}$, each of which gives an  exponentially small contribution respectively along backward and forward directions, therefore their multiplicative contribution becomes negligible. (As a side note, the contribution coming from the interference term vanishes upon performing the $\phi_k$ integration, as long as the net OAM satisfies: $\ell\neq 2$). These observations along with the exchange symmetry noted above help to significantly reduce the computation time and will  also  allow us to simplify the final expression for the photon yield, as we study specific cases in the next section.  

\section{Case Studies}
Here we study the scenarios in which $E$ and $\bar{E}$ have the following respective beam profiles: 1) $E=\text{LG}_{00},\, \bar{E}=\text{LG}_{00}$ where both beams are purely Gaussian,  2)  $E=\text{LG}_{00}+ \text{LG}_{10},\,  \bar{E}=\text{LG}_{00}$  where the first beam is in a mixture of the modes $\ell=0$ and $\ell=1$ with radial  index is set to zero  3) $E=\text{LG}_{00}+ \text{LG}_{10},\,  \bar{E}=\text{LG}_{00}+ \text{LG}_{10}$, where both beams are given as mixtures and finally 4)  $E=\text{LG}_{10},\, \bar{E}=\text{LG}_{00}$. The first case was considered in \cite{pe6}  for various collision angles. It will be particularly illustrative to analyze this case to benchmark our results with the earlier studies. As we will discuss in detail,  the second case comes with  the flip signature $\sim e^{-i \phi}$, whereas the  third case contains the signature  $\sim e^{2i \phi}$,  representing $\ell=2$ mode. From now on, for all the cases given above we set $\phi_{\ell\, p}=0$. Before we begin our analysis it is worth commenting on why we have specifically chosen  the pulses to be in a superposition state. The  flip  and doubling signatures in fact show up also in the case 4, which, from a technical perspective, is much simpler.  The corresponding amplitudes  in this case  are respectively given by (henceforth, we drop the $p$ index for convenience) $\bar{S}^{-++}_{1\,0\,0}$ and $ S^{++-}_{1\,1\,0} $. By the virtue of our earlier discussion, it is evident that the phase of these amplitudes oscillate rapidly therefore such transitions are exponentially suppressed. On the other hand, if one or both of the pulses are in a superposition,  the flip/doubling transitions as we shall see are respectively given by $|\bar{S}^{+-+}_{0\,1\,0}|^2$ and $|\bar{S}^{+-+}_{1\,0\,1}|^2$ which can constitute an appreciable signal. This in effect simulates the three-beam collision scenario discussed in \cite{orb}. Here, the beam given in superposition is basically two co-propagating beams which occupy different modes.  As the final case study, we shall nevertheless consider the case in which a pure Gauss mode, $\text{LG}_{00}$ and  $\text{LG}_{10}$ mode collide. The relevant signature in this case is given by $|\bar{S}^{+-+}_{1\,1\,0}|^2$, the angular spectrum of which encodes non-trivial topology. 

\subsection{Collision between pure Gaussian modes}

In view of the discussion of the previous section, we may write the signal photon density to a very good approximation  as 
\begin{eqnarray}
n(\vec{k})&\approx & 256\mathcal{P} \bigg(\sin^4 \frac{\theta_k}{2}  |\mathcal{S}(\vec{k})|^2 +\cos^4 \frac{\theta_k}{2}  |\mathcal{\bar{S}}(\vec{k})|^2  \bigg)
\label{pdens2} 
\end{eqnarray}
where we have neglected the interference term. In the remainder of this subsection we drop the $\ell$ index for convenience.  Recalling that $S^{+++}$ is exponentially suppressed and also the exchange symmetry: $S^{+-+}=S^{-++}$, from (\ref{ampsum}) we have: $|\mathcal{S}(\vec{k})|^2 \approx 1/16  | S^{+-+}|^2$. Similarly for  $\mathcal{\bar{S}}(\vec{k})$ we have  $\bar{S}^{+-+}=\bar{S}^{++-}$, which gives: $ |\mathcal{\bar{S}}(\vec{k})|^2 \approx  1/16 | \bar{S}^{+-+}|^2$. In the following we will use identical pulse parameters thus we set all the dimensionless ratios in (\ref{amp}) to unity.  Note that for this specific case  Fourier integrals in (\ref{famp}) have  reflection symmetry under $z \rightarrow -z, \, \theta_k \rightarrow \pi-\theta_k$,  such that  $S^{+-+} \rightarrow \bar{S}^{+-+}$. We will however represent the forward and backward emission amplitudes as separate for convenience. Performing the above simplifications and using (\ref{amp}) and (\ref{pdens2}) leads to
\begin{eqnarray}
N &\approx& \frac{2\alpha}{3\pi^2} \frac{1}{90^2}\frac{c^4 \tau^2}{\Omega^2}\left(\frac{z_R}{\lambda_c}\right)^4  \tilde{E}_0^6   \int  |\vec{k}|^3   e^{-\frac{\tau^2}{24}(|\vec{k}|c -\Omega )^2}  d|\vec{k}| \,  \nonumber\\
&\times & \bigg( \int  \, \sin{\theta_k} \sin^4 \frac{\theta_k}{2} \, d\theta_k    | \mathcal{F}^{+-+}|^2 +  \int  \, \sin{\theta_k} \cos^4 \frac{\theta_k}{2} \, d\theta_k    | \mathcal{\bar{F}}^{+-+}|^2 \bigg)
\label{yield1}
\end{eqnarray}
where we have denoted $\tilde{E}_0 =E_0/E_s$, and from now on we will use dimensionless field strengths normalized by $E_s$. The amplitudes $\mathcal{F}^{+-+}$ and $\mathcal{\bar{F}}^{+-+}$
 are given as ($\bar{z}=z/z_R$)
\begin{eqnarray}
\mathcal{F}^{+-+}&=&\int^{\infty}_{-\infty} d\bar{z}  \, \dfrac{e^{-\frac{32 \bar{z}^2 z_R^2}{3 c^2 \tau^2}} e^{-\frac{i}{3} \bar{z} z_R \big( (3\cos{\theta_k}-1)|\vec{k}| +4 \Omega/c \big) } e^{-\frac{i z_R |\vec{k}|^2 c  (1+\bar{z}^2)\sin^2{\theta_k}}{2\Omega (3i +\bar{z})}}}{3+ \bar{z}(2i +\bar{z})}\nonumber\\
\mathcal{\bar{F}}^{+-+}&=&\mathcal{F}^{+-+}(z \rightarrow -z, \, \theta_k \rightarrow \pi-\theta_k)
\label{ampz1}
\end{eqnarray}
where the substitutions in second line above are to be made at the integrand level. In order for the comparison with \cite{pe6}, we first note that mode dependent field strength is given as \cite{km2}
\begin{eqnarray}
\varepsilon_0 c \big (E_{0}^{\ell\, p} \big)^2\approx 8 \sqrt{\frac{2}{\pi}} \frac{\mathcal{E}}{\pi w_0^2 \tau} \frac{p!}{\big (p+ |\ell| \big )!}
\end{eqnarray}  
The pulse energy $\mathcal{E}$ here belongs to the specific mode $\{\ell, p\}$. It is specified w.r.t to the peak power at $z=0$ such that $\mathcal{E}=\varepsilon_0 c \int d t\, r dr\, d\phi \,|E(z=0)|^2$. Taking this into account and noting the relation: $z_R = \pi w^2_0/\lambda$, we fix the pulse parameters as follows: $\mathcal{E}=25$ J, $\lambda=800$ nm,  $\tau=25$ fs and $w_0= \lambda $. From (\ref{yield1}) we obtain the photon yield as $N=275$, which is in  excellent agreement with\cite{pe6}. Noting that the dominant contribution comes from the region $|\vec{k}|\sim \Omega $, we have specified the momentum integration interval here as  $(4\Omega/5, \, 6 \Omega/5)$. We did not observe any noticeable difference on the signal yield upon further increasing the integration range.  The number of signal photons coming from the signature  $S^{++-}$  is, as expected,  highly suppressed:  $N_b^{++-} \lessapprox 10^{-27}$  The yield coming from inelastic scattering  ($S^{+++}(|\vec{k}|\sim 3\Omega )$) is also  negligible:  $N_b^{+++}\approx 10^{-20}$.  By symmetry, the yield  for  all the signatures along the forward direction are identical to the ones coming from backward scattering. Note that in the rest of our  analysis we will focus our attention on the dominant signatures only.  

Before we move onto the next case, we estimate the signal yield based on  ELI-NP beam parameters \cite{ls2}. For each pulse we take: $\mathcal{E}=220$ J, $\lambda=800$ nm and  $\tau=25$ fs and set the focal radius close to its diffraction limited value: $w_0=2\, \mu$m.  The signal photon yield in this case becomes $N=7727 $.  As mentioned in the introduction, a large amount of these signal photons is emitted along the forward cones of the driving beams. It would be illustrative to estimate  the total number of the photons $N_{> \theta_d}$, that are emitted roughly outside the cones spanned by the beams. For this it convenient to define a mode-dependent angular distribution, which depends on the dominant signatures only. By using (\ref{ampsum}), (\ref{amp}) and (\ref{pdens2}) we may write 
\begin{eqnarray}
\rho_{\{\ell,\,p\}} &=& \frac{2\alpha}{\pi^2} \frac{1}{90^2}\frac{c^4 \tau^2}{\Omega^2}\left(\frac{z_R}{\lambda_c}\right)^4 \frac{\big(\tilde{E}_0^{\ell_1,\, p_1}\tilde{E}_0^{\ell_2,\, p_2}\tilde{E}_{\bar{0}}^{\bar{\ell}_3,\,\bar{p}_3} \big)^2}{\tau^2_{12}+2} \sin^4 \frac{\theta_k}{2} \sin{\theta_k} \nonumber\\&\times& \int_0^{\infty} |\vec{k}|^3  e^{-\frac{\tau^2}{16(\tau_{12}^2+2)} (|\vec{k}|c-\Omega_{21}\Omega)^2} | \mathcal{F}^{+ - +}_{\{\ell,\,p\}}|^2 d|\vec{k}|\nonumber\\
\bar{\rho}_{\{\ell,\,p\}}&=& \frac{2\alpha}{\pi^2} \frac{1}{90^2}\frac{c^4 \tau^2}{\Omega^2}\left(\frac{z_R}{\lambda_c}\right)^4 \frac{\big(\tilde{E}_0^{\ell_1,\, p_1}\tilde{E}_{\bar{0}}^{\bar{\ell}_2,\, \bar{p}_2}\tilde{E}_{\bar{0}}^{\bar{\ell}_3,\,\bar{p}_3} \big)^2}{2\tau^2_{12}+1} \cos^4 \frac{\theta_k}{2} \sin{\theta_k} \nonumber\\&\times& \int_0^{\infty} |\vec{k}|^3  e^{-\frac{\tau^2}{16(2\tau_{12}^2+1)} (|\vec{k}|c-\Omega)^2} | \mathcal{\bar{F}}^{+-+}_{\{\ell,\,p\}}|^2 d|\vec{k}|
\label{angspec}
\end{eqnarray}  
\begin{figure}
\includegraphics[width=10 cm,height=5.5cm]{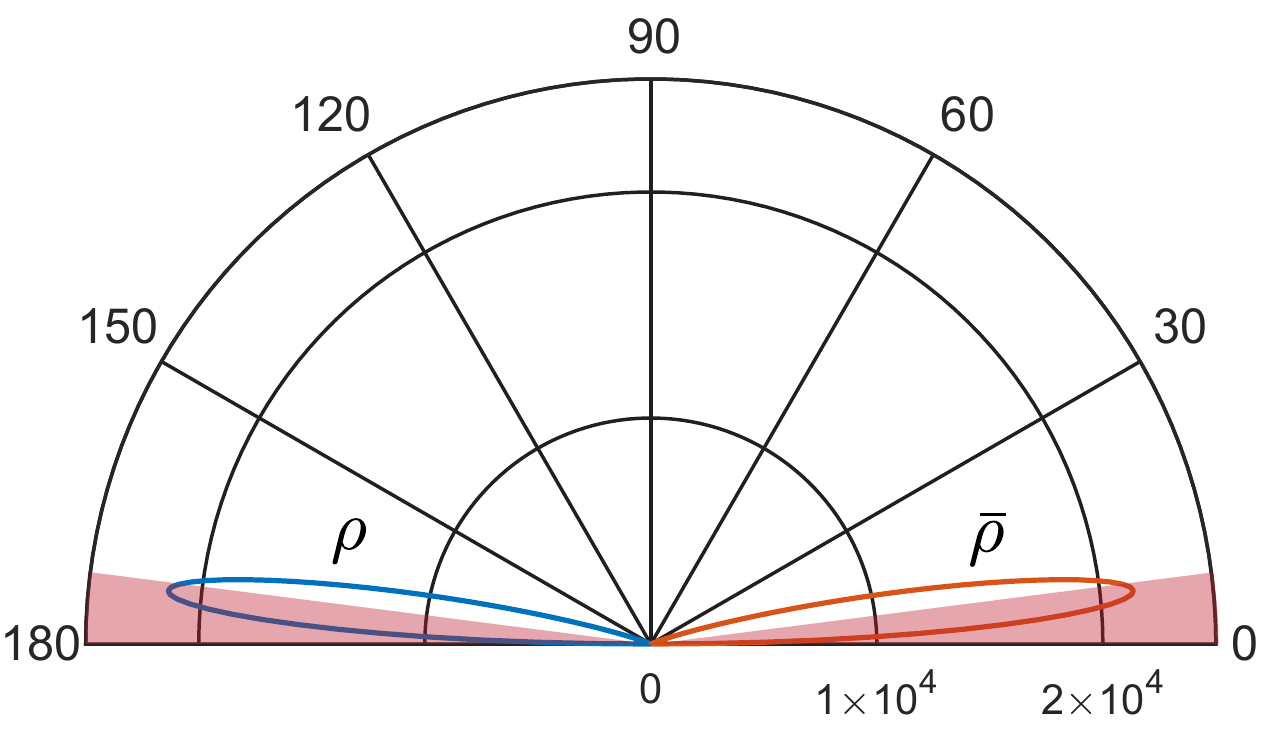}
\caption{Angular spectrum of signal photon for backward (blue) and forward (orange) scattering along the optical axis $z$. The divergence of the external beams are represented by the shaded (red) cones. Pulse parameters are given  as: $\tilde{E}_0=\tilde{E}_{\bar{0}}=9.80648\times 10^{-4},\, w_0= 2\, \mu\text{m},\, \tau=25 \, \text{fs}$ and  $\lambda= 800\, \text{nm}$, yielding $N_{> \theta_d}=3925$ (explained in the text). Note that we have dropped the subscripts $\{\ell,\, p\}$.} 
\label{f1}
\end{figure}Note that (\ref{angspec}) by definition includes only the diagonal terms in $|\mathcal{S}(\vec{k})|^2$  and $|\mathcal{\bar{S}}(\vec{k})|^2 $ for which $\phi_k$ integration has been carried out, yielding an overall $2\pi$ factor at the end. We have also included the multiplicity factor of 4 coming from the exchange symmetry.  In general, any  $\phi_k$ dependence brought by the cross terms in $|\mathcal{S}(\vec{k})|^2$  and $|\mathcal{\bar{S}}(\vec{k})|^2 $ is of the form: $e^{i m \phi_k}$,  therefore  such terms do not contribute to the photon number as long as $m$ is an integer. On the other hand the cross terms that appear with  vanishing total OAM  must be separately accounted for.  Using the above definitions we define : $N^f_{\{\ell,p\}\, >\theta_d}=\int^{\pi}_{\theta_d } \bar{\rho}_{\{\ell,p\}}\, d\theta_k $ for the forward emission and likewise: $N^b_{\{\ell,p\},\, >\theta_d}=\int^{\pi-\theta_d}_{0} \rho_{\{\ell,p\}}\, d\theta_k $ for the backward emission. Here,  $\theta_d$ is the divergence half-angle of the beam given by the relation $\theta_d=M^2\lambda/\pi w_0$, where $M$ is the beam quality factor (this is also referred as  beam propagation factor). $M^2$ is defined as unity for the ideal case of Gaussian beam \cite{siegman}. For higher order LG modes quality factor is defined as\cite{ss}: $M^2=2 p + |\ell|+1$.  Setting $M^2=1$, the total number of such signal photons based on ELI-NP beam parameters is $N_{> \theta_d}=3925$. The angular spectrum of the signal photons is shown in Figure \ref{f1}. We would like to draw attention to the importance of parameter $w_0$ here. Recalling that signal scales as $\sim \tilde{E}^6_0$, one could expect having smaller values of waist size could prove advantageous because of the higher field strength but on the other hand smaller waist size results in larger divergence of the beam. This leads to a lesser percentage of the signal photons that are emitted outside the forward cones of the external beams. Using the parameters of \cite{pe6}, we find that $N_{> \theta_d}=112$, which corresponds to $\% 40$ of the total signal yield whereas using ELI-NP parameters; with the same pulse duration and wavelength but with larger waist size, this becomes $\% 50$.  A similar observation has been recently noted in \cite{gkk}.
\subsection{Collision between LG modes}

\begin{figure}
\includegraphics[width=8 cm,height=5 cm]{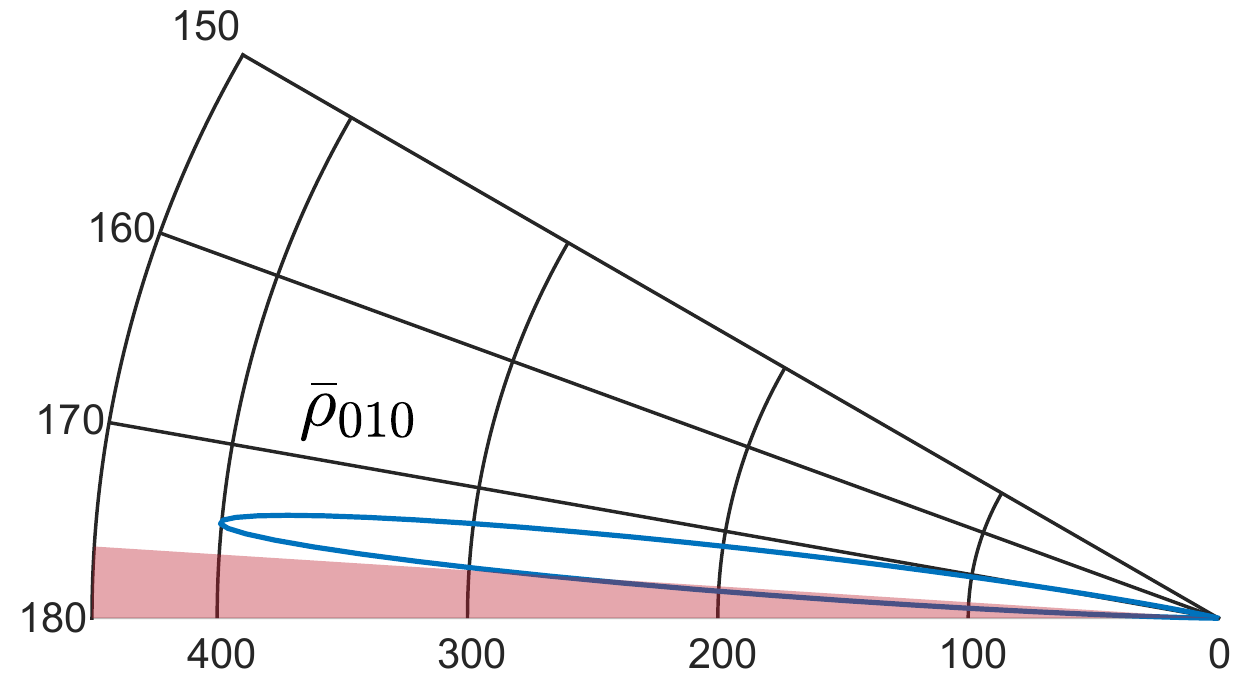}
\caption{Angular spectrum of signal photon for backward scattering along the optical axis $z$. Pulse parameters are given as: $\tilde{E}^0_0=4.90324 \times 10^{-4},\, \tilde{E}^1_0=4.33042 \times 10^{-4},\,\tilde{E}^0_0=5.37123\times 10^{-4},\,  w_0= 2\, \mu\text{m},\, \tau=25 \, \text{fs}$ and  $\lambda= 800\, \text{nm}$ (counter-propagating pulse is frequency doubled), yielding $N^b_{010,\, > \theta_d}=38$. Note that we have dropped the subscripts $\{p\}$.} 
\label{f3}
\end{figure}

Here  we first  consider the collision between two pulses, first of which is prepared in a superposition of LG$_{00}$ and LG$_{10}$ modes and the latter is prepared in LG$_{00}$.     Using (11) photon densities  $|\mathcal{S}(\vec{k})|^2$  and $|\mathcal{\bar{S}}(\vec{k})|^2 $ are now given as
\begin{eqnarray}
|\mathcal{S}(\vec{k})|^2&\approx &\frac{1}{16}\big|    S^{+ - +}_{0\,0\,0} 
+  S^{+ - +}_{1\,0\,0}  +  S^{+ - +}_{0\,1\,0} +  S^{+ - +}_{1\,1\,0}\big|^2\nonumber\\
|\mathcal{\bar{S}}(\vec{k})|^2 &\approx &\frac{1}{16}\big| \bar{S}^{+ - +}_{0\,0\,0}
 +   \bar{S}^{+ - +}_{1\,0\,0}  \big|^2
 \label{ampcase2}
\end{eqnarray}
As before, we have used the exchange symmetry defined in (\ref{exsym}) and have neglected all the suppressed terms. The specific signature that we are interested in is given by $S^{+ - +}_{0\,1\,0}$. Using  (\ref{amp}), it can be written as
\begin{eqnarray}
S^{+- +}_{0\,1\,0}  &=& e^{-i  \phi_k} \tilde{E}_0^{0}\tilde{E}_0^{1}\tilde{E}_{\bar{0}}^{0}  \bigg(\frac{2\pi z^2_{R} c}{\Omega}\bigg) \bigg(\frac{\pi^\frac{1}{2}\tau}{2\sqrt{\tau^2_{12}+2}}e^{-\frac{\tau^2}{16(\tau_{12}^2+2)} (|\vec{k}|c-\Omega_{21}\Omega)^2} \bigg)    \mathcal{F}^{+-+}_{0\,1\,0}
\end{eqnarray}    
where
\begin{eqnarray}
\mathcal{F}^{+-+}_{0\,1\,0}&=&|\vec{k}| \sqrt{\frac{c z_R}{\Omega}}z^{21}_{R}\sin{\theta_k} \int^{\infty}_{-\infty} d\bar{z}  \, e^{-\delta_0} e^{i (\delta_1+ \delta_2 )} \dfrac{(\bar{z}+i)(i\bar{z}+z^{21}_{R})}{\big( 2 z^{21}_R+\Omega_{21}+\bar{z} (2i+\bar{z}\Omega_{21})\big)^2}\nonumber\\
\delta_0&=&\frac{32 \bar{z}^2 z_R^2}{c^2 \tau^2(1+2/\tau^2_{12})}\nonumber\\
\delta_1&=&  \bar{z}z_{R}\bigg( \frac{c |\vec{k}| (\tau^2_{12}-2)+4\Omega \Omega_{21}}{c (2+\tau^2_{12})} + |\vec{k}|\cos{\theta_k} \bigg)\nonumber\\
\delta_2&=& i\frac{c |\vec{k}|^2 z_{R} (1+\bar{z}^2)(i\bar{z}+z^{21}_{R})\sin^2{\theta_k}}{2\Omega \big( 2 z^{21}_R+\Omega_{21}+\bar{z} (2i+\bar{z}\Omega_{21})\big)}
\end{eqnarray}    
\begin{figure}
\includegraphics[width=14 cm,height=8 cm]{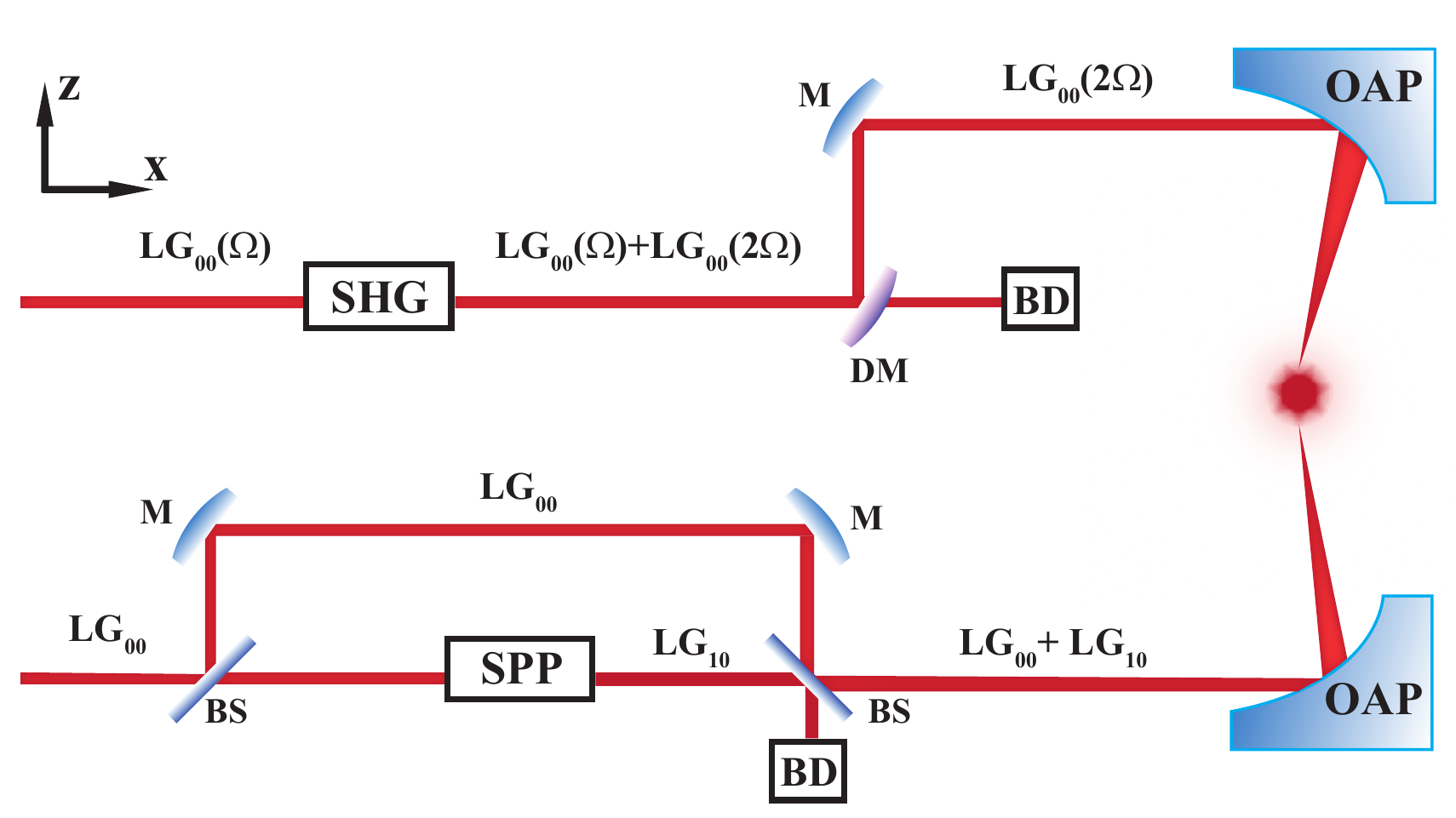}
\caption{The schematics of the proposed experimental setup probing the OAM flip signature. On the lower arm, one of the 2×10 Petawatt laser arms at ELI-NP is split by 50:50 beam splitter (BS). The transmitted beam carrying $\%50$ of the source intensity is sent to Spiral Phase Plate (SPP), which converts the incoming LG$_{00}$ mode into LG$_{10}$ mode with $\%78$ conversion efficiency. The output LG$_{10}$ is combined with LG$_{00}$ coming from the mirror (M) via 50:50 BS at a $\%50$  loss in intensity. The resultant beam is sent to the off-axis parabolic mirror for focusing (OAP). The upper arm is frequency double via second harmonic generation with $\%30$  conversion efficiency. The resulting frequency doubled modes are picked by a dichroic mirror (DM) and sent to OAP for focusing. The remaining modes are directed to a beam dump (BD).} 
\label{f2}
\end{figure}
In order to probe the flip signature, we envision a scenario in which one of the 2$\times$10 PW arms at ELI-NP is split into two. The transmitted beam carrying $\%50$ of the source intensity is directed to a large diameter  spiral phase plate (SPP), tailored for the intense, short-pulse applications \cite{spp1,spp2}.  With $\%$78 SPP conversion efficiency,  the output  LG$_{10}$ mode is recombined with LG$_{0,0}$.  The pulse from the second arm is frequency doubled with a conversion efficiency $\%$30.  The schematics of the would-be experimental setup  is depicted in Figure \ref{f2}.  We use the same set of parameters of the previous section for  LG$_{00}$ mode.  We set $\Omega_{21}=z^{21}_{R}=2$ and  $\tau_{12}=1$. Taking into account the efficiency factors (see Figure \ref{f2}) we find $N_{010}  =38 $ signal photons emitted with $\ell=-1$.  Using (\ref{angspec}) and the definition of  $N^b_{\{\ell,p\},\, >\theta_d}$, we also find  that virtually all of the signal photons are emitted outside forward cone of the (frequency-doubled) beam propagating along $-z$. The angular spectrum  is shown in Figure \ref{f3}.  

In order to  estimate the yield for the doubling signature we consider the case where both beams are  prepared in a superposition of  LG$_{00}$ and  LG$_{10}$ modes.  In view of the scenario outlined above,  this case entails the split-recombination procedure on the both 10 PW arms.  From here on out, our treatment basically follows the same steps. For the backward emission, the dominant signature that we are interested is 
\begin{eqnarray}
S^{+- +}_{1\,0\,1}  &=& e^{2 i  \phi_k} \tilde{E}_0^{1}\tilde{E}_0^{0}\tilde{E}_{\bar{0}}^{1}  \bigg(\frac{2\pi z^2_{R} c}{\Omega}\bigg) \bigg(\frac{\pi^\frac{1}{2}\tau}{2\sqrt{\tau^2_{12}+2}}e^{-\frac{\tau^2}{16(\tau_{12}^2+2)} (|\vec{k}|c-\Omega_{21}\Omega)^2} \bigg)    \mathcal{F}^{+-+}_{1\,0\,1}
\end{eqnarray}    
where 
\begin{eqnarray}
\mathcal{F}^{+-+}_{1\,0\,1}&=&\frac{ |\vec{k}|^2  c z_R}{\Omega} z^{21}_{R}(z^{21}_{R}\Omega_{21})^{1/2}\sin^2{\theta_k}  \int^{\infty}_{-\infty} d\bar{z}  \, e^{-\delta_0} e^{i (\delta_1+ \delta_2 )} \dfrac{(\bar{z}-i)(\bar{z}-i z^{21}_{R}) (1 +\bar{z}^2)}{\big( 2 z^{21}_R+\Omega_{21}+\bar{z} (2i+\bar{z}\Omega_{21})\big)^3}
\end{eqnarray}   
and for the forward emission we have
\begin{eqnarray}
\bar{S}^{+- +}_{1\,0\,1}  &=& e^{2 i  \phi_k} \tilde{E}_0^{1}\tilde{E}_{\bar{0}}^{0}\tilde{E}_{\bar{0}}^{1}  \bigg(\frac{2\pi z^2_{R} c}{\Omega}\bigg) \bigg(\frac{\pi^\frac{1}{2}\tau}{\sqrt{2\tau^2_{12}+1}}e^{-\frac{\tau^2}{16(2\tau_{12}^2+1)} (|\vec{k}|c-\Omega)^2} \bigg)    \mathcal{\bar{F}}^{+-+}_{1\,0\,1}
\end{eqnarray}    
where
\begin{eqnarray}
\mathcal{\bar{F}}^{+-+}_{1\,0\,1}&=&\frac{ |\vec{k}|^2  c z_R}{\Omega}(z^{21}_{R} \Omega_{21})^{1/2} (z^{21}_{R} \sin {\theta_k})^2  \int^{\infty}_{-\infty} d\bar{z}  \, e^{-\bar{\delta}_0} e^{i (\bar{\delta}_1 +\bar{\delta}_2) } 
\dfrac{(i +  \bar{z})(\bar{z}- i z^{21}_{R})\big(\bar{z}+ i z^{21}_{R} \big)^{2}}{\big( \bar{z} (\bar{z} -2 i z^{21}_{R}\Omega_{21})+z^{21}_{R} (z^{21}_{R}+2\Omega_{21})\big)^3}\nonumber\\
\bar{\delta}_0&=&\frac{32 \bar{z}^2 z_R^2}{c^2 \tau^2(2+1/\tau^2_{12})}\nonumber\\
\bar{\delta}_1&=&  \bar{z}z_{R}\bigg( \frac{c |\vec{k}| (2\tau^2_{12}-1)-4\Omega \tau^2_{12} }{c (2\tau^2_{12}+1)} + |\vec{k}|\cos{\theta_k} \bigg)\nonumber\\
\bar{\delta}_2&=& i\frac{c |\vec{k}|^2 z_{R} (1-i\bar{z})(\bar{z}^2 +(z^{21}_{R})^2)\sin^2{\theta_k}}{2\Omega \big( \bar{z} (\bar{z} -2 i z^{21}_{R}\Omega_{21})+z^{21}_{R} (z^{21}_{R}+2\Omega_{21})\big)}
\end{eqnarray}
\begin{figure}
\includegraphics[width=10 cm,height=5.5 cm]{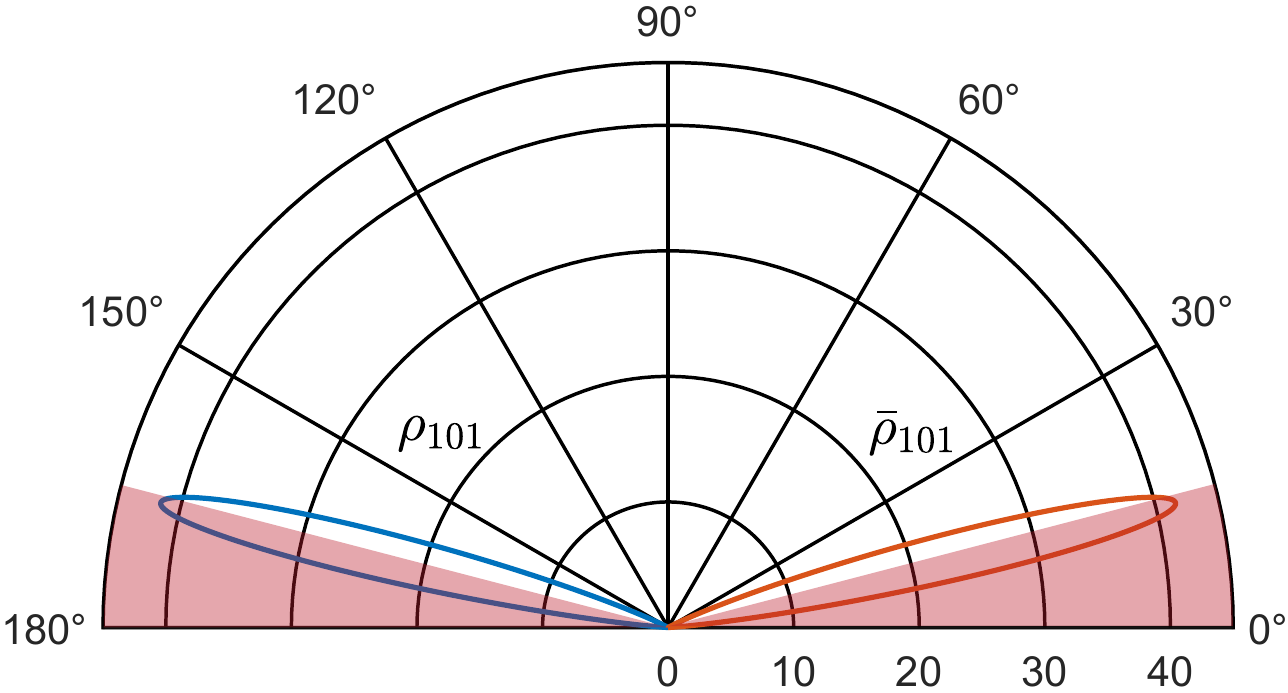}
\caption{Angular spectrum of signal photon for backward and forward scattering along the optical axis $z$. Pulse parameters are given as: $\tilde{E}^1_0=\tilde{E}^1_{\bar{0}}=4.33042 \times 10^{-4},\, \tilde{E}^0_0=\tilde{E}^0_{\bar{0}}=4.90324 \times 10^{-4},\,  w_0= 2\, \mu\text{m},\, \tau=25 \, \text{fs}$ and  $\lambda= 800\, \text{nm}$,  yielding $N_{> \theta_d}=7 $. } 
\label{f4}
\end{figure}Using previously given pulse parameters and the conversion efficiencies,  we find $N_{101}=15$ signal photons, 7 of which are emitted outside the forward cone of  LG$_{10}$ mode (see Figure \ref{f4}).  

As our final example we look into the collision between pure  $\text{LG}_{10}$ (propagating along $+\hat{z}$) and  $\text{LG}_{00}$ modes.  Backward emission amplitude is given by $|\mathcal{S}(\vec{k})|^2 \approx 1/16 \big|    S^{+ - +}_{1\,1\,0} \big|^2$. Note that this signature is already contained within (\ref{ampcase2})  but we will nevertheless treat this as a separate and much simpler collision scenario, in which one of the 10 PW arms remains unaltered and the latter is directly sent to SPP.  The  amplitude $S^{+- +}_{1\,1\,0} $ can be written as
\begin{eqnarray}
S^{+- +}_{1\,1\,0}  &=& \big(\tilde{E}_0^{1}\big)^2 \tilde{E}_{\bar{0}}^{0}  \bigg(\frac{2\pi z^2_{R} c}{\Omega}\bigg) \bigg(\frac{\pi^\frac{1}{2}\tau}{2\sqrt{\tau^2_{12}+2}}e^{-\frac{\tau^2}{16(\tau_{12}^2+2)} (|\vec{k}|c-\Omega_{21}\Omega)^2} \bigg)    \mathcal{F}^{+-+}_{1\,1\,0}
\end{eqnarray}    
where
\begin{eqnarray}
\mathcal{F}^{+-+}_{1\,1\,0}&=&   \int^{\infty}_{-\infty} d\bar{z}  \, e^{-\delta_0} e^{i (\delta_1+ \delta_2 )} \dfrac{2\big(i\bar{z}+z^{21}_{R}\big)^3 (1+ i\delta_2 )}{\big( 2 z^{21}_R+\Omega_{21}+\bar{z} (2i+\bar{z}\Omega_{21})\big)^2}
\label{ampzcase4}
\end{eqnarray}

\begin{figure}
\includegraphics[width=10 cm,height=6 cm]{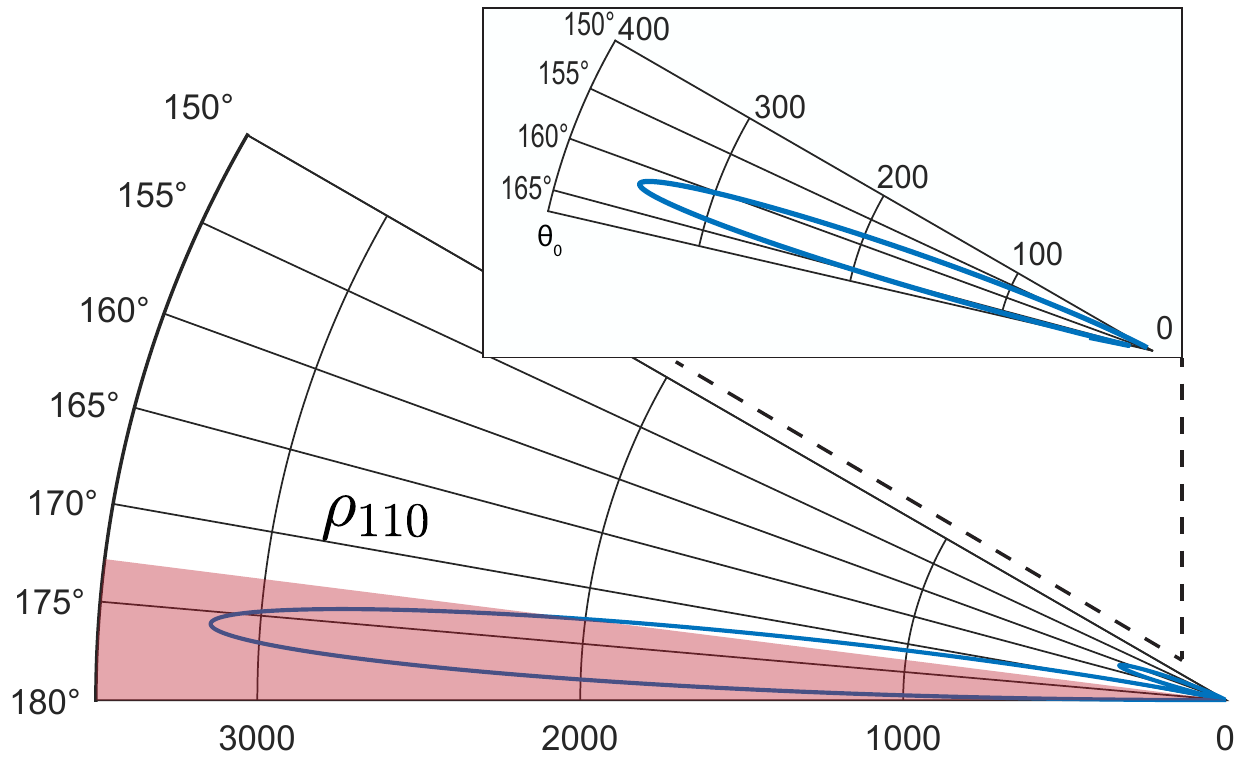}
\caption{Angular spectrum of signal photon for  backward scattering along the optical axis $z$. Pulse parameters are given as: $\tilde{E}^1_0=8.66084 \times 10^{-4},\, \tilde{E}^0_{\bar{0}}=9.80648\times 10^{-4} ,\,  w_0= 2\, \mu\text{m},\, \tau=25 \, \text{fs}$ and  $\lambda= 800\, \text{nm}$,  yielding $N^{b}_{110,\, >\theta_d}=107$ and $N^{b}_{110,\, >\theta_0}=46 $ } 
\label{f5}
\end{figure}The distinguishing property of this signature is that  the integrand in (\ref{ampzcase4}) contains a non-trivial zero whose location is given by $\delta_2(\theta_0)=i$.  Assuming equal pulse parameters ($z^{21}_{R}=\Omega_{21}=\tau_{12}=1$), and letting  $|\vec{k}|\rightarrow\Omega/c$, at the pulse center (z=0)  we have $\theta_0=\pi -\arcsin{\sqrt{6 c/z_R \Omega}}$. Further setting $w_0= 2\, \mu\text{m},\, \tau=25 \, \text{fs}$ and  $\lambda= 800\, \text{nm}$ as before we find $\theta_0\approx \ang{167.26}$,  which accurately marks the location of the minimum shown in the angular spectrum in Figure \ref{f5}. Around this minimum,  the spectrum displays a multi-lob structure with two distinguished peaks for the signal. This kind of structure happens to be the property of the signatures, which encode a reduction in the topological charge;  here in this specific case the external photons with non-vanishing OAM ($\ell=1$) scatter into a signal state with vanishing OAM. The precise condition for the general case is
\begin{eqnarray}
|\ell_1| + |\ell_2| +|\bar{\ell}_3| > s_1 \ell_1 + s_2 \ell_2 + s_3 \bar{\ell_3}
\label{cond}
\end{eqnarray}
which follows from the behavior of the Hypergeometric function (see Appendix for the details). We find the total signal yield  as $N_{110}=401$, $N^{b}_{110,\, >\theta_d}=107$ and $N^{b}_{110,\, >\theta_0}=46 $.

\section{Conclusions and Outlook}

In this study we have elaborated on the analytic structure of the photon emission amplitudes, induced by counter-propagating, pulse-shaped LG beams of arbitrary mode decomposition. We have found that the emission amplitude for the signal yield can greatly be simplified by neglecting the rapidly oscillating components of the amplitude and by making use of the exchange symmetry. We have identified nonlinear OAM  signatures for three distinct cases and presented our estimates for the signal photon yield for each case. We should emphasize that the assumption of head-on collision that we have used, although may not fully reflect the experimental conditions,  has ultimately given us the semi-analytic formulas, which we believe can provide useful estimates and insights into the emission spectrum. It would be therefore desirable to extend such analysis to the scenarios that include arbitrary collision angle between the beams.  The broadband structure of the pulse is another important aspect which can be further refined by the considering the flat-top or super-Gaussian profiles. This could  also be relevant for the production of LG modes, either via SPP or spiral phase mirror (SPM) \cite{spm}, whose step height must match the multiples of the central wavelength.  Depending on the broadband character of the seed beam,  SPP (SPM) can induce additional modes of non-integer OAM \cite{frac}, which may require further investigation.

\section*{Acknowledgements}

We thank Felix Karbstein, Keita Seto and Takahisa Jitsuno for fruitful discussions and comments. This work was carried out under the Nucleu contract PN 19 06 01 05 funded by the Ministry of Research, Innovation and Digitalization.

\section{Appendix}
Here we give the details on the master formula for the emission amplitude, specific cases of which were given in section III.  In the following we will work out the steps that goes into the calculation of $\mathcal{S}(\vec{k})$.  Recalling (\ref{famp}) and (\ref{ef}),  we write $\mathcal{S}(\vec{k})$ explicitly as
\begin{eqnarray}
\mathcal{S}(\vec{k})&=&\frac{1}{8} \sum_{\substack{\ell_1,\, \ell_2,\,\bar{\ell}_3,\\ p_1,\,p_2,\,\bar{p}_3 }} \sum_{\substack{s_1 \, s_2 \,s_3}} \int d^4 x \, e^{i k x} \mathcal{E}_{\,\ell_1\,\ell_2 \,\bar{\ell}_3}^{p_1\,p_2\,\bar{p}_3} \, \text{Exp}\left[i\Phi_{s_1\,s_2\,s_3} \right]
\end{eqnarray}
Note that 
\begin{eqnarray}
kx=-\omega t+ |\vec{k}| r \sin{\theta_k} \cos{\big( \phi-\phi_k \big)}+  |\vec{k}| z \cos{\theta_k} 
\end{eqnarray}
and
\begin{eqnarray}
\mathcal{E}_{\,\ell_1\,\ell_2 \,\bar{\ell}_3}^{p_1\,p_2\,\bar{p}_3}&=& E_0^{\ell_1 \, p_1}E_0^{\ell_2\, p_2}E_{\bar{0}}^{\bar{\ell}_3\,\bar{p}_3}  e^{-8(z/c-t)^2/\tau^2}e^{-4(t+z/c)^2/\bar{\tau}^2}e^{-\frac{2 r^2 z_{R}^2}{w_0^2 (z^2 +z_{R}^2)}} e^{-\frac{ r^2 \bar{z}_{R}^2}{\bar{w}_{0}^2 (z^2 +\bar{z}_{R}^2)}}\nonumber\\
&\times& \frac{z^2_{R}}{z^2 + z^2_{R}}\frac{\bar{z}_{R}}{(z^2 + \bar{z}^2_{R})^{1/2}} \left(\frac{\sqrt{2} r z_{R}}{w_0\sqrt{z^2 +z^2_{R}}}\right)^{|\ell_1|+|\ell_2|}  \left(\frac{\sqrt{2} r \bar{z}_{R}}{\bar{w}_0\sqrt{z^2 +\bar{z}^2_{R}}}\right)^{|\bar{\ell}_3|}   \nonumber\\ 
 &\times &  L^{|\ell_1|}_{p_1}\left(\frac{2 r^2}{w(z)^2}\right) L^{|\ell_2|}_{p_2}\left(\frac{2 r^2}{w(z)^2}\right) L^{|\bar{\ell}_3|}_{\bar{p}_3}\left(\frac{2 r^2}{\bar{w}(z)^2}\right) \nonumber\\
\Phi_{s_1\,s_2\,s_3}&=&  (s_1 + s_2 + s_3\Omega_{21})\Omega \, t - (s_1 + s_2 - s_3\Omega_{21}) \Omega \frac{z}{c}  + (s_1 \ell_1 + s_2 \ell_2 + s_3 \bar{\ell}_3)\phi \nonumber\\
&-&(s_1 \phi_{\ell_1 \, p_1} + s_2 \phi_{\ell_2 \, p_2} + s_3 \phi_{\bar{\ell}_3 \, \bar{p}_3}) - (s_1 + s_2)\frac{z}{z_{R}}\frac{r^2}{w(z)^2}+s_3\frac{z}{\bar{z}_{R}}\frac{r^2}{\bar{w}(z)^2}\nonumber\\ 
&+&  \big(2(s_1 p_1 + s_2 p_2 )+ (s_1 |\ell_1| + s_2 |\ell_2| ) + (s_1 + s_2 ) \big)\text{tan}^{-1}\frac{z}{z_{R}} -s_3(2 \bar{p}_3 + |\bar{\ell}_3| +1) \text{tan}^{-1}\frac{z}{\bar{z}_{R}}\nonumber\\
\end{eqnarray}
Isolating the time-dependent terms, Gaussian integral over time can be readily performed, giving
\begin{eqnarray}
\mathcal{S}_{t}(\vec{k}) &=&\int_{-\infty}^{\infty} dt\, e^{i(s_1+s_2+s_3\Omega_{21})\Omega t}e^{-8 (z/c-t)^2/\tau^2}e^{- 4 (z/c+t)^2/\bar{\tau}^2}e^{-i\omega t}\nonumber\\
&=&e^{-\frac{32z^2}{c^2 (\tau^2 +2\bar{\tau}^2)}}e^{iz \frac{\tau^2-2\bar{\tau}^2}{c(\tau^2+2\bar{\tau}^2)}(\omega-(s_1+s_2+s_3\Omega_{21})\Omega)}e^{-\frac{\tau^2\bar{\tau}^2}{16 (\tau^2+2\bar{\tau}^2)} (\omega-(s_1+s_2+s_3\Omega_{21})\Omega)^2}\frac{\sqrt{\pi}}{2\sqrt{\frac{2}{\tau^2}+\frac{1}{\bar{\tau}^2}}}
\end{eqnarray}
which upon reorganizing  the terms reads
\begin{eqnarray}
\mathcal{S}_{t}(\vec{k}) &=&\bigg(\frac{\pi^\frac{1}{2}\tau}{2\sqrt{2+\tau^2_{12}}}e^{-\frac{\tau^2}{16(\tau_{12}^2+2)} (|\vec{k}|c-(s_1+s_2+s_3\Omega_{21})\Omega)^2} \bigg) e^{-\frac{32 \bar{z}^2 z_R^2}{c^2 \tau^2(1+2/\tau^2_{12})}}e^{iz \frac{\tau_{12}^2-2}{(\tau^2_{12}+2)}(|\vec{k}|-(s_1+s_2+s_3\Omega_{21})\Omega/c)}\nonumber
\end{eqnarray}
The angular integral is given as 
\begin{eqnarray}
\mathcal{S}_{\phi}(\vec{k}) &=&\int_0^{2\pi} d\phi \, e^{i (s_1 \ell_1 +s_2 \ell_2 +s_3 \bar{\ell}_3)\phi} e^{i |\vec{k}| r \sin\theta_k \cos(\phi-\phi_k)}\nonumber\\
&=& 2\pi  e^{i(s_1 \ell_1 +s_2 \ell_2 +s_3 \bar{\ell}_3)\phi_{k}} \sum_{n=0}\frac{(-1)^{n+\frac{1}{2}(s_1 \ell_1 +s_2 \ell_2 +s_3 \bar{\ell}_3)}}{n!\Gamma[n+s_1 \ell_1 +s_2 \ell_2 +s_3 \bar{\ell}_3+1]} \left(\frac{\vec{k}r|\sin{\theta_{k}}|}{2}\right)^{2n + s_1 \ell_1 +s_2 \ell_2 +s_3 \bar{\ell}_3}
\end{eqnarray}
where we have made use of the representation
\begin{equation}
\int_0^{2\pi} \text{exp}\left[i\ell \phi + i k b \rho \cos{(\phi-\theta+\pi)} \right]=2\pi e^{-i\frac{\pi}{2}\ell}J_{\ell}(k b \rho)e^{i\ell \theta}
\end{equation}
and performed series expansion of the Bessel function. Expanding Laguerre polynomials and collecting the $r$-dependent terms, the radial integral can be performed in a similar fashion by making use of the first formula of 3.462 in \cite{gr}, ultimately giving (henceforth we use the shorthand notation: $b=s_1 \ell_1 +s_2 \ell_2 +s_3 \bar{\ell}_3 +1$)
\begin{eqnarray}
\mathcal{S}_{r}(\vec{k}) &=&\int_0^{\infty} dr\,  e^{-\frac{2 r^2 z_{R}^2}{w_0^2 (z^2 +z_{R}^2)}
-\frac{r^2 \bar{z}_{R}^2}{\bar{w}_0^2 (z^2 +\bar{z}_{R}^2)}}
e^{-i(s_1 +s_2)\frac{r^2 z z_{R}}{w_0^2 (z^2 + z_{R}^2)}+is_3\frac{r^2 z \bar{z}_{R}}{\bar{w}_0^2 (z^2 + \bar{z}_{R}^2)}}\nonumber\\
&\times& \left(\frac{ r^2 z_{R}^2 }{w_0^2 (z^2 +z_{R}^2)}\right)^{i_1+i_2 +\frac{1}{2} (|\ell_1|+|\ell_2|)}
\left(\frac{ r^2 \bar{z}_{R}^2 }{\bar{w}_0^2 (z^2 +\bar{z}_{R}^2)}\right)^{\bar{i}_3+\frac{1}{2} |\bar{\ell}_3|}  r^{2n +b}\nonumber\\
&=& \left(\frac{z_{R}\Omega}{2c(z^2+z^2_{R})}\right)^{i_1+i_2 +\frac{1}{2}(|\ell_1|+|\ell_2|)}\left(\frac{\bar{z}_{R}\bar{\Omega}}{2c(z^2+\bar{z}^2_{R})}\right)^{\bar{i}_3 +\frac{1}{2}|\bar{\ell}_3|}
\nonumber\\
&\times & \frac{1}{2}\left(\frac{(i(s_1+s_2)z+2z_{R})\Omega}{2c(z^2+z^2_{R})} +\frac{(-is_3z +\bar{z}_{R})\bar{\Omega}}{2c(z^2+\bar{z}^2_{R})}\right)^{-(a+n)}\Gamma[a+n]
\end{eqnarray}
where $a=1+i_1+i_2+\bar{i}_3+\frac{1}{2} (s_1 \ell_1 + s_2 \ell_2 +s_3 \bar{\ell}_3 +|\ell_1| + |\ell_2| + |\bar{\ell}_3|) $ and we  have used the definition: $z_R=\pi w_0^2/\lambda $. The summation over the index $n$ can be done by using the series expansion of the Hypergeometric function yielding
\begin{eqnarray}
\sum_{\substack{\,n =0}} &\ & \left(|\vec{k}| \sin{\theta_{k}} \right)^{2n } \frac{(-1)^{n}\Gamma\left[a+n\right] }{ 2^{2n}\, n!\, \Gamma[b + n] }\bigg(\frac{(i (s_1+s_2) z+2z_{R})\Omega}{2c(z^2+z^2_{R})} +\frac{(-is_3 z +\bar{z}_{R})\bar{\Omega}}{2c(z^2+\bar{z}^2_{R})} \bigg)^{-n} \nonumber\\ 
&=&\Gamma\left[ a \right] {}_{1}{\tilde{F}}_{1}\left[a,\,b ,\, d \right] 
\end{eqnarray}
where all the factors above, except $n!$ and Gamma functions, are encoded in $d$, which is given by($\bar{z}=z/z_R$)
\begin{eqnarray}
d &=&  \frac{i c |\vec{k}|^2 z_{R}(\bar{z}^2+1)\big(\bar{z}^2+\big(z^{21}_{R}\big)^2\big) \sin^2{\theta_k}}
{2\big ((s_1+s_2)\bar{z}-2i\big )\big(\bar{z}^2+\big(z^{21}_{R}\big)^2\big)\Omega -2(\bar{z}^2+1)(s_3 \bar{z}+iz^{21}_{R})\Omega_{21}\Omega}
\end{eqnarray}
In the final step we have used the definition of the regularized Hypergeometric function: ${}_{1}{\tilde{F}}_{1}\left[a, \, b, \, d \right]=\lim_{x \to b}{}_{1}{F}_{1}\left[a, \, x, \,d \right]/\Gamma[x]$.
The remaining factors coming from $\mathcal{S}_{r}(\vec{k}) $ can be rewritten as
\begin{eqnarray}
&&\bigg(\frac{(i (s_1+s_2) z+2 z_{R})\Omega}{2c(z^2+z^2_{R})} +\frac{(-is_3 z + \bar{z}_{R})\bar{\Omega}}{2c(z^2+\bar{z}^2_{R})} \bigg)^{-a}\left(\frac{z_{R}\Omega}{2c(z^2+z^2_{R})}\right)^{i_1 + i_2+ \frac{1}{2}(|\ell_1|+|\ell_2|)}\nonumber\\
&\times&
\left(\frac{\bar{z}_{R}\bar{\Omega}}{2c(z^2+\bar{z}^2_{R})}\right)^{\bar{i}_3 +\frac{1}{2}|\bar{\ell}_3|}
\nonumber\\
&=&\bigg(\frac{i (s_1 +s_2)\bar{z}+2}{\bar{z}^2+1} +\frac{(-i s_3 \bar{z} + z^{21}_{R})\Omega_{21}}{\bar{z}^2+\big(z^{21}_{R}\big)^2} \bigg)^{-a}\nonumber\\
&\times&\left(\frac{1}{\bar{z}^2+1}\right)^{i_1 + i_2+ \frac{1}{2}(|\ell_1|+|\ell_2|)}\left(\frac{z^{21}_{R}\Omega_{21}}{ \bar{z}^2+\big(z^{21}_{R}\big)^2}\right)^{\bar{i}_3 +\frac{1}{2}|\bar{\ell}_3|}\left(\dfrac{\Omega}{2 c z_{R}}\right)^{-\big(1 +1/2(b-1)\big)}
\end{eqnarray}
Written this way, all the dimensionful quantities are contained in the last factor above. This last factor can be combined with the term coming from $\mathcal{S}_{\phi}(\vec{k}) $ such that
\begin{eqnarray}
\left(\dfrac{\Omega}{ c z_{R}}\right)^{-\big(1 +1/2(b-1)\big)}\left(|\vec{k}| \sin{\theta_{k}} \right)^{ b-1}=\left(|\vec{k}| \left(\frac{c z_{R}}{\Omega}\right)^{1/2} \sin{\theta_{k}} \right)^{ b-1} \frac{c z_{R}}{\Omega}
\end{eqnarray}
where the first term remains dimensionless. Rewriting the phase terms by using the relation
\begin{eqnarray}
e^{i s \arctan \bar{z}}=\left(\frac{1+i\bar{z}}{1-i\bar{z}}\right)^{\frac{s}{2}}, \quad s \in \mathbb{R}
\end{eqnarray}
and collecting all the remaining factors and recalling (11),  the final result is given as
\begin{eqnarray}
S^{s_1 s_2 s_3}_{\{\ell,\, p \}}  &=& E_0^{\ell_1\, p_1}E_{0}^{\ell_2\, p_2}E_{\bar{0}}^{\bar{\ell}_3\,\bar{p}_3}  \bigg(\frac{2\pi z^2_{R} c}{\Omega}\bigg) \bigg(\frac{\pi^\frac{1}{2}\tau}{2\sqrt{\tau^2_{12}+2}}e^{-\frac{\tau^2}{16(\tau_{12}^2+2)} \big(|\vec{k}|c-(s_1+s_2+s_3\Omega_{21})\Omega\big)^2} \bigg)   \nonumber\\
&\times & e^{i (s_1 \ell_1 + s_2\ell_2 +s_3 \bar{\ell}_3) \phi_k} \mathcal{F}^{s_1 s_2 s_3}_{\{\ell,\, p \}}
\end{eqnarray}
where
\begin{eqnarray}
 \mathcal{F}^{s_1 s_2 s_3}_{\{\ell,\, p \}} &=&e^{-i (s_1 \phi_{\ell_1\, p_1} + s_2 \phi_{\ell_2\, p_2} +s_3  \phi_{\bar{\ell}_3\, \bar{p}_3})} \sum_{i_1,\, i_2,\, \bar{i}_3=0}^{p_1,\, p_2,\, \bar{p}_3}\binom{p_1 + |\ell_1|}{p_1-i_1}\binom{p_2 + |\ell_2|}{p_2-i_2}\binom{\bar{p}_3 + |\bar{\ell}_3|}{\bar{p}_3-\bar{i}_3} \nonumber\\ &\times& \left(|\vec{k}| \left(\frac{c z_{R}}{\Omega}\right)^{1/2} \sin{\theta_{k}} \right)^{ b-1} \frac{(-1)^{i_1 +  i_2 + \bar{i}_3 +\frac{1}{2}(b-1)}}{i_1! i_2! \bar{i}_3!}2^{a-b}\Gamma\left[a\right] \nonumber\\ 
&\times & \int_{-\infty}^{\infty} d \bar{z}\,  e^{-\frac{32 \bar{z}^2 z^2_{R}}{c^2 \tau^2(1+2/\tau_{12}^2)}} e^{i\bar{z}z_{R} \big(|\vec{k}|\cos\theta_{k}-(s_1+s_2-s_3\Omega_{21})\Omega/c\big)} e^{i\bar{z} z_{R} \frac{\tau_{12}^2-2}{\tau_{12}^2+2}\big(|\vec{k}|-(s_1+s_2+s_3\Omega_{21})\Omega/c\big)}\nonumber\\
&\times& \left(\frac{1+i\bar{z}}{1-i\bar{z}}\right)^{\frac{1}{2}\big(2(s_1 p_1 + s_2 p_2 )+ (s_1 |\ell_1| + s_2 |\ell_2| ) + s_1 + s_2\big) }\left(\frac{1+i\bar{z}/z^{21}_{R}}{1-i\bar{z}/z^{21}_{R}}\right)^{-\frac{1}{2}s_3(2 \bar{p}_3+ |\bar{\ell}_3|  +1)}\nonumber\\ 
&\times &\left(\frac{1}{\bar{z}^2+1}\right)^{i_1 + i_2+ \frac{1}{2}(|\ell_1|+|\ell_2|)}\left(\frac{z^{21}_{R}\Omega_{21}}{ \bar{z}^2+ \big (z^{21}_{R}\big)^2}\right)^{\bar{i}_3 +\frac{1}{2}|\bar{\ell}_3|}\frac{1}{\bar{z}^2 + 1}\frac{z^{21}_{R}}{\big(\bar{z}^2 + \big (z^{21}_{R}\big)^2\big)^{1/2}}\nonumber\\
&\times&\bigg(\frac{i (s_1 +s_2)\bar{z}+2}{\bar{z}^2+1} +\frac{(-i s_3 \bar{z} + z^{21}_{R})\Omega_{21}}{\bar{z}^2+\big (z^{21}_{R}\big)^2} \bigg)^{-a} {}_{1}{\tilde{F}}_{1}\left[a,\,b,\, d \right]
\end{eqnarray}
Note that we have made the substitution $z=\bar{z} z_{R}$.  The hypergeometric function appearing above can be rewritten  as \cite{as}
\begin{eqnarray}
{}_{1}{\tilde{F}}_{1}\left[a,\,b,\,d \right]&=&\lim_{x \to a}\bigg( \frac{\Gamma[1-x]}{\Gamma[b-x]} \bigg) L_{-a}^{b-1}(d) \nonumber\\
&=& \lim_{x \to a}\bigg( \frac{\Gamma[1-x]}{\Gamma[b-x]} \bigg)  L_{a-b}^{b-1}(-d) e^{d}
\end{eqnarray}
The second line above can be verified upon series expansion and using the reflection property: $L_{-a}(d)=e^{\alpha} L_{a-1}(-d)$.  An important consequence of the above relation is that when $a>b$, the amplitude develops non-trivial zeroes whose location coincide the with the zeroes of  $ L_{a-b}^{b-1}(-d) $. Using the definitions of $a$ and $b$, it is easy to see that this occurs when:
\begin{eqnarray}
i_1+i_2+\bar{i}_3+\frac{1}{2} (|\ell_1| + |\ell_2| + |\bar{\ell}_3|-s_1 \ell_1 - s_2 \ell_2 - s_3\bar{\ell}_3  ) >0 
\end{eqnarray} 
Setting the radial modes to zero, this precisely  becomes the condition given in (\ref{cond}). The same arguments can be given for the forward emission amplitude, the derivation of which can be performed by making use of  similar steps. For brevity, here we simply give the final result:
\begin{eqnarray}
\bar{S}^{s_1 s_2 s_3}_{\{\ell,\, p \}}  &=& E_0^{\ell_1,\, p_1}E_{\bar{0}}^{\bar{\ell}_2,\, \bar{p}_2}E_{\bar{0}}^{\bar{\ell}_3,\,\bar{p}_3}  \bigg(\frac{2\pi z^2_{R} c}{\Omega}\bigg) \bigg(\frac{\pi^\frac{1}{2}\tau}{2\sqrt{2\tau^2_{12}+1}}e^{-\frac{\tau^2}{16(2\tau_{12}^2+1)} \big(|\vec{k}|c-(s_1+(s_2+s_3)\Omega_{21})\Omega\big)^2} \bigg)   \nonumber\\
&\times & e^{i (s_1 \ell_1 + s_2\bar{\ell}_2 +s_3 \bar{\ell}_3) \phi_k} \mathcal{\bar{F}}^{s_1 s_2 s_3}_{\{\ell,\, p \}}
\end{eqnarray}
with
\begin{eqnarray}
 \mathcal{\bar{F}}^{s_1 s_2 s_3}_{\{\ell,\, p \}} &=&e^{-i (s_1 \phi_{\ell_1 \, p_1} + s_2 \phi_{\bar{\ell}_2 \, \bar{p}_2} +s_3  \phi_{\bar{\ell}_3 \, \bar{p}_3})} \sum_{i_1,\, \bar{i}_2,\, \bar{i}_3=0}^{p_1,\, \bar{p}_2,\, \bar{p}_3}\binom{p_1 + |\ell_1|}{p_1-i_1}\binom{\bar{p}_2 + |\bar{\ell}_2|}{\bar{p}_2-\bar{i}_2}\binom{\bar{p}_3 + |\bar{\ell}_3|}{\bar{p}_3-\bar{i}_3} \nonumber\\ &\times& \left(|\vec{k}| \left(\frac{c z_{R}}{\Omega}\right)^{1/2} \sin{\theta_{k}} \right)^{ \bar{b}-1} \frac{(-1)^{i_1 +  \bar{i}_2 + \bar{i}_3 +\frac{1}{2}(\bar{b}-1)}}{i_1!\bar{i}_2! \bar{i}_3!}2^{\bar{a}-\bar{b}}\Gamma\left[\bar{a}\right] \nonumber\\ 
&\times & \int_{-\infty}^{\infty} d \bar{z}\,  e^{-\frac{32 \bar{z}^2 z^2_{R}}{c^2 \tau^2(2+1/\tau_{12}^2)}} e^{i\bar{z}z_{R} \big(|\vec{k}|\cos\theta_{k}-(s_1-(s_2+s_3)\Omega_{21})\Omega/c\big)} e^{i\bar{z} z_{R} \frac{2\tau_{12}^2-1}{2\tau_{12}^2+1}\big(|\vec{k}|-(s_1+(s_2+s_3)\Omega_{21})\Omega/c\big)}\nonumber\\
&\times& \left(\frac{1+i\bar{z}}{1-i\bar{z}}\right)^{\frac{1}{2} s_1 ( 2 p_1 +  |\ell_1|  +1) }\left(\frac{1+i\bar{z}/z^{21}_{R}}{1-i\bar{z}/z^{21}_{R}}\right)^{-\frac{1}{2}\big(2 (s_2 \bar{p}_2+s_3 \bar{p}_3)+ (s_2 |\bar{\ell}_2|+s_3 |\bar{\ell}_3|)   +s_2 + s_3 )\big)}\nonumber\\ 
&\times &\left(\frac{1}{\bar{z}^2+1}\right)^{i_1 + \frac{1}{2} |\ell_1|}\left(\frac{z^{21}_{R}\Omega_{21}}{ \bar{z}^2+ \big (z^{21}_{R}\big)^2}\right)^{\bar{i}_2+\bar{i}_3 +\frac{1}{2}(|\bar{\ell}_2|+|\bar{\ell}_3|)}\frac{1}{\big(\bar{z}^2 + 1\big)^{1/2}}\frac{\big(z^{21}_{R}\big)^2}{\big(\bar{z}^2 + \big (z^{21}_{R}\big)^2\big)}\nonumber\\
&\times&\bigg(\frac{i s_1 \bar{z}+1}{\bar{z}^2+1} +\frac{(-i (s_2 + s_3 )\bar{z} + 2 z^{21}_{R})\Omega_{21}}{\bar{z}^2+\big (z^{21}_{R}\big)^2} \bigg)^{-\bar{a}} {}_{1}{\tilde{F}}_{1}\left[\bar{a},\,\bar{b},\, \bar{d} \right]
\end{eqnarray}
where we have defined
\begin{eqnarray}
\bar{a}&=&1+i_1+\bar{i}_2+\bar{i}_3+\frac{1}{2} (s_1 \ell_1 + s_2 \bar{\ell_2} +s_3 \bar{\ell}_3 +|\ell_1| + |\bar{\ell}_2| + |\bar{\ell}_3|)\nonumber\\
\bar{b}&=& 1+  s_1 \ell_1 + s_2 \bar{\ell_2} +s_3 \bar{\ell}_3\nonumber\\
\bar{d}&=& \frac{i c |\vec{k}|^2 z_{R} (\bar{z}^2+1)(\bar{z}^2+\big(z^{21}_{R}\big)^2) \sin^2{\theta_k}}
{2(s_1\bar{z}-i)(\bar{z}^2+\big(z^{21}_{R}\big)^2)\Omega-2(\bar{z}^2+1)((s_2+s_3) \bar{z}+2iz^{21}_{R})\Omega_{21}\Omega}
\end{eqnarray}

\end{document}